\input epsf
\font\upright=cmu10 scaled\magstep1
\font\sans=cmss10
\newcommand{\be}{\begin{equation}}
\newcommand{\ee}{\end{equation}}
\newcommand{\bea}{\begin{eqnarray}}
\newcommand{\eea}{\end{eqnarray}}
\newcommand{\ssf}{\sans}
\newcommand{\Z}{\hbox{\upright\rlap{\ssf Z}\kern 2.7pt {\ssf Z}}}
\newcommand{\R}{\hbox{\upright\rlap{I}\kern 1.7pt R}}
\newcommand{\PP}{\hbox{\upright\rlap{I}\kern 1.8pt P}}
\newcommand{\bI}{{\bf I}}
\newcommand{\bJ}{{\bf J}}
\newcommand{\tbJ}{{{\tilde\bJ}}}
\newcommand{\bL}{{\bf L}}
\newcommand{\vr}{{\vec{r}}}
\newcommand{\dA}{{\dot{A}}}
\newcommand{\vtau}{{\vec{\tau}}}
\newcommand{\p}{{\partial}}
\newcommand{\tr}{{\rm tr}}

\documentstyle[12pt]{article}

\begin{document}
\title{
\vskip 30pt
{\bf \LARGE \bf Semiclassical Quantization of Hopf Solitons}} 
\vskip 3.0 cm

\author{Wang-Chang Su \\[10pt]
\\{\normalsize{\sl Department of Physics}}
\\{\normalsize{\sl National Chung-Cheng University}}
\\{\normalsize{\sl Chia-Yi, 621, Taiwan}}
\\
\\{\normalsize{\sl Email : suw@phy.ccu.edu.tw}}\\ 
}
\maketitle
\vskip 25pt

\begin{abstract}

The gauge equivalent formulation of the Faddeev-Skyrme model is used for the study of the quantum theory.
The rotational quantum excitations around the soliton solution of Hopf number unity are investigated by the method of collective coordinates.
The quantum Hamiltonian of the system is found to coincide with the Hamiltonian of a symmetrical top rotating in $SU(2)$.     
Thus, the irreducible representations of physical observables can be constructed.   

\end{abstract} 
\newpage
 
Among the topological field theories in (3+1)-dimensional space-time, perhaps the Faddeev-Skyrme model \cite{faddeev} is the best-known theory admitting localized topological solitons \cite{KR}, in which the soliton number does not arise as a winding number.
The Lagrangian of the model consists of a scalar field whose target space is $S^2$. 
In order for a solution to have finite energy, the scalar field has to be fixed at spatial infinity.
The Euclidean $\R^3$ space is then regarded as a compactified $S^3$ space. 
Hence, at any fixed time, the scalar field defines a map from $S^3$ space to $S^2$ target.
Such mapping falls into nontrivial homotopy classes, \( \pi_3 (S^2) = \Z \), whose topological charge is called the Hopf number.  
For a given field configuration, the Hopf number represents the linking number between two different field lines. 

The topology of the Faddeev-Skyrme model has received intensive study in the past few years \cite{numerical,GH,analytical,su}; especially by numerical simulations, thanks to the advances in computer power \cite{numerical,GH}. 
The numerical results have revealed the intricate and fascinating topological structures of the model. 
However, most of the investigations mainly focus on the search for classical static solitons with nonzero Hopf number.
Less attention has been paid to the discussion of quantum theory of the system \cite{GH}.
Yet, in order to determine the physical observables and to find a representation of those observables on a Hilbert space, it is equally important to investigate the quantum structure of the model. 

The present letter is aimed at complementing those reports on the classical solutions.
Here, the leading contribution of quantum excitations around the classical ground state is examined.
More precisely, only rotational symmetry modes of the model are treated quantum mechanically.
To demonstrate this, the gauge equivalent formulation of the Faddeev-Skyrme model \cite{su,baal}, instead of the original Lagrangian, is used for the study of the quantum theory.
 By introducing collective coordinates for the quantization of rotational modes, we derive the quantum Hamiltonian. 
It is found that the Hamiltonian at quantum level coincides with the Hamiltonian of a symmetrical top rotating in $SU(2)$.
Thus, the construction of a basis of quantum states that spans the Hilbert space can be easily achieved.   

We start with the gauge equivalent version of the Faddeev-Skyrme model. 
The identification of this version with the original Faddeev-Skyrme model is explained in \cite{su}.
In that paper, the solitonic solutions of this gauge equivalent Lagrangian were analyzed by the rational map ansatz. 
A simple expression of energy function was found and the Bogomolny-type lower bound of energy was established.
Explicitly, the gauge equivalent Lagrangian takes the form  
\be
L \, = \, 4 \int d^3x \, 
\left[ 
L^a_\mu \, L^a_\mu 
\, - \, \left( \epsilon^{3ab} \, L^a_\mu \, L^b_\nu \right)^2 
\right] \, .
\label{FS} 
\ee
In this equation, $L^a_\mu$ for $a = 1,2$ represents the first and second components of an $SU(2)$ Lie-algebra valued current \( \bL_\mu \). 
The current is defined by 
\be 
\bL_\mu \, \equiv \, i \, \tau_i \, L_\mu^i \, = \, U^{-1} \p_\mu U \, , 
\label{bL}
\ee
where the index \( i = 1,2,3 \) and the scalar field $U(\vr, t)$ is a generic element in $SU(2)$ Lie group.
Note that the Lagrangian (\ref{FS}) is invariant under a residual $U(1)$ gauge transformation of $SU(2)$, which is $U \to Uh$, where $h = \exp(i \tau^3 \alpha/2)$.
Therefore, the gauge invariant (physical) fields take values in the coset space 
\( SU(2)/U(1) = S^2 \).

For static field configurations, the energy function can be found by using the rational map ansatz \cite{su}. 
We restrict ourselves to the discussion of Hopfions here, that is, the soliton solutions of Hopf number unity.
In this case, the scalar field $U_0 (\vr)$ in (\ref{bL}) is expressed in terms of the usual hedgehog solution
\be
U_0 (\vr) 
\, = \, 
\exp \left( \, i f(r) \, \frac{\vr}{r} \cdot \vtau \, \right) \, ,
\label{ansatz}
\ee
where the subscript $0$ denotes that of the static hedgehog solution. 
$\vtau = ( \tau_1 , \tau_2 , \tau_3 )$ are the Pauli matrices and the vector \( \vr = (x,y,z) \) gives a direction in the $SU(2)$ Lie-algebra. 
$f(r)$ is a radial profile function, which satisfies the boundary conditions \( f(0) = \pi \) at the origin and $f(\infty) = 0$ at spatial infinity.
Thus, the profile function $f(r)$ is a monotonically decreasing function.

Using (\ref{bL}) and (\ref{ansatz}), it can be shown after angular integration that the static energy function of the Hopfion configuration is given as \cite{su}
\be
E_0 
\, = \,
\frac{32}{3} \, \pi \int dr \,
\Bigg[ \>
r^2 f^{\, 2}_r  
\, + \, 
2 \, \left( 1 + f^{\, 2}_r \right) \, \sin^2 \! f 
\, + \,
\frac{\sin^4 \! f}{r^2} \, 
\Bigg] \, ,
\label{QH1}
\ee
where \( f_r = df/dr \).
The minimum $M_0$ of this energy function has to be numerically estimated by the variation of the profile function $f(r)$.
The result is \( M_0 = 1.232 \times 32 \pi^2 \).
Analytically, if the kink profile function \cite{sutcliffe} \( f(r) = 4 \tan^{-1} [ \, \exp(-r) \, ] \) is used instead of the numerical calculated profile function, the result becomes \( M_0 = 1.240 \times 32 \pi^2 \) that gives a good approximation.
Note that, up to an over-all normalization constant, the Hopfion static energy (\ref{QH1}) is identical to that of the standard hedgehog solution of the Skyrme model \cite{HMS}.

Now, we turn to the quantization of the hedgehog soliton (\ref{ansatz}) by applying the collective coordinate method \cite{ANW}.
For a thorough review on similar treatment of the Skyrme model, please refer to \cite{books} and references therein.
The central idea of this semiclassical method is to merely quantize the quantum fluctuations that are generated by symmetries of the Lagrangian (\ref{FS}).
In other words, it provides the leading contribution to quantum excitations around the classical solution.
The standard procedure is to take the time-dependent $SU(2)$ matrix $A(t)$ as the collective coordinates which describe the spin (isospin) degrees of freedom.
Explicitly, we have the following expression
\be
U (\vr,t) \, = \, A(t) \, U_0 (\vr) \, A^{-1} (t) \, .
\label{ansatz2}
\ee

Next, we insert (\ref{ansatz2}) into the Faddeev-Skyrme Lagrangian (\ref{FS}) and integrate out the degrees of freedom other than the collective coordinates $A(t)$.
This leads to the effective Lagrangian as a function of collective coordinates and velocities $\dA (t)$: 
\be
L(A, \dA) \, = \, 
\, - \, M_0
\, - \, 2
\left[ \,
(I_1 + I_2) \left( \dA A^{-1} \right)_{\! i} \! \left( \dA A^{-1} \right)_{\! i}
\, + \,
(I_1 - I_2) \left( \dA A^{-1} \right)_{\! 3} \! \left( \dA A^{-1} \right)_{\! 3} \,
\right] \, , 
\label{lagrangian}
\ee
where the subscript $i$ takes values from 1 to 3.
$M_0$ is the classical Hopfion mass (\ref{QH1}).
$\dA$ means $dA/dt$ and \( \left( \dA A^{-1} \right)_{\! i} \) denotes the $i$-th component of the matrix
\be
\dA A^{-1} \, = \, \tau_i \left( \dA A^{-1} \right)_{\! i}
\, = \, \frac{\tau_i}{2} \, \tr \left( \tau_i \, \dA A^{-1} \right) \, .
\ee 
In the effective Lagrangian (\ref{lagrangian}), $I_1$ and $I_2$ are called the moments of inertia for the rotating Hopfion, whose physical meaning will be apparent when the Hamiltonian is obtained. They are given by
\bea
I_1 & = &
\frac{2 \pi}{15} \int dr \, 
\Bigg[
\left( 5 \sin^2 \! f \, \cos^2 \! f + \sin^4 \! f \right) \, + \,
2 \, \frac{\sin^4 \! f}{r^2} \, + \,
8 f^{\, 2}_r \, \sin^2 \! f \, \cos^2 \! f  \, 
\Bigg] \, ,
\label{I1}
\\
I_2 & = &
\frac{2 \pi}{15} \int dr \, 
\Bigg[ \,
8 \sin^4 \! f \, + \,
6 \, \frac{\sin^4 \! f}{r^2} \, + \,
f^{\, 2}_r \, 
\left( 10 \sin^4 \! f - 6 \sin^2 \! f \, \cos^2 \! f \right) 
\Bigg] \, .
\label{I2}
\eea 
Both integrals are then numerically calculated using the same kink profile function, which is mentioned above in the paragraph right after the energy function (\ref{QH1}).
Their values are approximately \( I_1 \approx 4.192 \) and \( I_2 \approx 12.282 \).

Observe that the effective Lagrangian (\ref{lagrangian}) is invariant under two kinds of transformation on the collective coordinates $A(t)$. 
The first transformation is the left multiplication \( A \to A W \), where \( W = e^{i \, \tau_i \theta_i/2} \) is a generic $SU(2)$ matrix. 
This corresponds to an arbitrary three-space rotation.
It is the symmetry of the Lagrangian (\ref{lagrangian}) because of the isotropy of space. 
The other one is the right multiplication \( A \to V A \), where \( V = e^{i \, \tau_3 \theta_3/2} \) lies in the $U(1)$ subgroup of $SU(2)$.
This symmetry is associated with the isospin rotation of the system in the $\tau_3$ direction.  
The left and right conserved currents of the corresponding symmetry are obtained from the usual Neother construction,
\bea
\bJ_i & = &
- 2i \, \Bigg[ \,
(I_1 + I_2) \left( \dA A^{-1} \right)_{\! j} \! R_{j i} \, + \,
(I_1 - I_2) \left( \dA A^{-1} \right)_{\! 3} \! R_{3 \, i} \, \Bigg] \, ,
\label{bJi}
\\
\bI_3 & = &
- 4i \, I_1 \left( \dA A^{-1} \right)_{\! 3} \, .
\label{bI3}
\eea  
Normally, the left current $\bJ_i$ is called spin angular momentum and the right one $\bI_3$ is called generator about the third isospin axis, respectively. 
The Lagrangian (\ref{lagrangian}) thus has the \( U_L (1) \times SU_R (2) \) symmetry.
$R_{i j}$ in (\ref{bJi}) is the matrix in the adjoint representation of $SU(2)$ group, which is defined through the following relation
\be
A \, \tau_i \, A^{-1} \, = \, \tau_j \, R_{j i} (A) \, .
\ee 

Since \( A(t) \in SU(2) \), one can locally parameterize the collective coordinates by three independent generalized ``coordinates'' $q_i$, for $i$ = 1 to 3.
Then, the generalized ``momenta'' conjugate to $q_i$ are defined as \( p_i = \p L/\p {\dot q}_i \). 
Here $L$ is the effective Lagrangian (\ref{lagrangian}). 
By definition $q_i$ and $p_i$ satisfy the canonical Poisson brackets: \( \{ q_i,q_j \} = \{ p_i,p_j \} = 0 \) and \( \{ q_i,p_j \} = \delta_{ij} \). 
Therefore, one is able to deduce the Poisson brackets among all three components of spin angular momentum and the third isospin generator, which are \( \{ \bJ_i, \bJ_j \} =  \epsilon_{ijk} \, \bJ_k \) and \( \{ \bJ_i, \bI_3 \} = 0 \). 

The Hamiltonian of the Faddeev-Skyrme system in terms of the generators (\ref{bJi}) and (\ref{bI3}) is derivable from the well-known Hamilton's principle, i.e., \( H (q,p) = p_i \, {\dot q}_i - L (q,{\dot q}) \).
The result is found to be
\be
H \, = \, M_0 
\, + \, \frac{1}{2 (I_1 + I_2)} \, \bJ_i \, \bJ_i 
\, - \, \frac{I_1 - I_2}{4 I_1 (I_1 + I_2)} \, \bI_3 \, \bI_3 \, .
\label{hamiltonian}
\ee 
Further, we can reshuffle the above Hamiltonian into a more elucidative form by introducing a rotated set of angular momentum \( \tbJ_i \equiv R_{i j} \, \bJ_j \).
In this definition, we notice that \( \tbJ_i \tbJ_i = \bJ_i \bJ_i \) and \( \tbJ_3 = \bI_3 \).
The physics of introducing the rotated angular momentum is to align the third component of spin angular momentum in the direction of the third isospin generator.   
As a result, the Hamiltonian (\ref{hamiltonian}) takes the simple expression
\be  
H \, = \, M_0
\, + \, \frac{1}{2 (I_1 + I_2)} \, 
\left( \, \tbJ_1 \, \tbJ_1 \, + \, \tbJ_2 \, \tbJ_2 \, \right)
\, + \, \frac{1}{2 \cdot 2 I_1} \, \tbJ_3 \, \tbJ_3 \, .
\label{hamiltonian1}
\ee 
This (\ref{hamiltonian1}) precisely describes a symmetrical top rigidly rotating in $SU(2)$.
It is the Hamiltonian for an axially symmetric rotator.
The inertia tensor of this system is distributed as follows; two of three principal moments of inertia are equal to $(I_1 + I_2)$ but different from the third one that has value $2I_1$.

To proceed to the quantum theory of the Hamiltonian (\ref{hamiltonian}), we follow the standard Dirac's quantization prescription.
The procedures consist of promoting the classical variables $\bJ_i$ and $\bI_3$ to quantum operators and replacing the Poisson brackets among them by commutation relations.  
Afterwards, we need to find an irreducible representation of those commutators on the Hilbert space \cite{landau}. 
Denote the $(2j+1)$-dimensional irreducible representation of the rotation group by the matrix \( D^{\, j} (A) \), which depends on the collective coordinates $A(t)$.
Because \( \{ H, \bJ^2, \bJ_3, \bI_3 \} \) form a complete set of mutually commuting operators, the energy eigenfunctions are given by the matrix element \( D^{\, j}_{k \, m} (A) \) of the rotational matrix \( D^{\, j} (A) \).
In the notation \( D^{\,j}_{k \, m} (A) \), the index $k$ is the eigenvalue of the third isospin generator $\bI_3$ and the index $m$ is the eigenvalue of the third component of the angular momentum $\bJ_3$.
The orientation of the $\bI_3$-axis with respect to the $\bJ_3$-axis defines the eigenvalues $k$ and $m$. 

As a matter of fact, higher dimensional representations of the rotational group can be established by tensor products of the collective coordinates $A(t)$, because the matrix $A$ itself is the fundamental spinor representation of $SO(3)$ group.
However, there are two inequivalent classes of irreducible representations.
The first class has representations with even polynomials in the matrix $A(t)$, which correspond to tensorial wave functions with integer spin.
The second one has those with odd polynomials in $A(t)$, which correspond to spinorial wave functions with a half-odd-integer spin. 
 
In conclusion, the mass formula of the spinning Hopfion in the state \( D^{\, j}_{k \, m} (A) \) is 
\be
M \, = \, M_0 
\, + \, \frac{j(j+1)}{2 (I_1 + I_2)} 
\, - \, \frac{k^2 (I_1 - I_2)}{4 I_1 (I_1 + I_2)} \, ,
\label{energy}
\ee
which is independent of the eigenvalue $m$, since the top is invariant under any rotation of the coordinate axes. 
The extraction of predictions for physical observables in this system is performed similar to that done in the Skyrme model.    
For example, the expectation value of a quantum operator in the Hilbert space is expressed as a volume integral over the $SU(2)$ group.
Furthermore, the matrix elements of physical observables can be calculated in the like manners.

\section*{Acknowledgment}
\noindent 
This work was supported in part by Taiwan's National Science Council Grant No. 90-2112-M-194-005.

\end{document}